\documentclass[twoside,11pt]{article}

\usepackage[accepted]{melba}

\usepackage{amsmath,amsfonts}

\usepackage{tabularx,adjustbox,booktabs}
\usepackage[super]{nth}
\usepackage{adjustbox}
\usepackage{tabularx}
\usepackage{subcaption}
\usepackage{ulem}

\newcommand{\norm}[1]{\left\lVert#1\right\rVert}

\newcommand\clearrow{\global\let\rowmac\relax}

\melbaid{2023:001}  
\melbaauthors{Yoon \textit{et~al}}
\volume{2}
\firstpageno{1}  
\year{2023}  
\datesubmitted{11/2022} 
\datepublished{01/2023}  

\ShortHeadings{Automated Cardiac Resting Phase Detection}{Yoon \textit{et~al.}}

\title{Automated Cardiac Resting Phase Detection Targeted on the Right Coronary Artery}

\author{\firstname Adrian V. \surname Dalca \email adalca@mit.edu \\  
	\addr EECS, Massachusetts Institute of Technology, Cambridge, MA, USA
	\AND
	\name Mert R. Sabuncu\orcid{0000-1111-2222-3333} \email msabuncu@cornell.edu \\
	\addr School of Electrical and Computer Engineering, Cornell University, Ithaca, NY, USA
}
\author{\firstname Seung Su \surname Yoon \email seung.su.yoon@fau.de \\  
	\addr Pattern Recognition Lab, Department of Computer Science, Friedrich-Alexander University Erlangen-Nuremberg, Erlangen 91058, Germany \\
	\addr Magnetic Resonance, Siemens Healthcare GmbH, Erlangen 91052, Germany
	\AND
	\firstname Elisabeth \surname Preuhs \email elisabeth.preuhs@gmail.com \\
	\addr Pattern Recognition Lab, Department of Computer Science, Friedrich-Alexander University Erlangen-Nuremberg, Erlangen 91058, Germany
	\AND
	\firstname Michaela \surname Schmidt \email michaela.schmidt@siemens-healthineers.com \\
	\addr Magnetic Resonance, Siemens Healthcare GmbH, Erlangen 91052, Germany
	\AND
	\firstname Christoph \surname Forman \email christoph.forman@siemens-healthineers.com \\
	\addr Magnetic Resonance, Siemens Healthcare GmbH, Erlangen 91052, Germany
	\AND
	\firstname Teodora \surname Chitiboi \email teodora.chitiboi@siemens-healthineers.com \\
	\addr Siemens Medical Solutions USA, Inc., Princeton, NJ, USA
	\AND
	\firstname Puneet \surname Sharma \email sharma.puneet@siemens-healthineers.com \\
	\addr Siemens Medical Solutions USA, Inc., Princeton, NJ, USA
	\AND
	\firstname Juliano L. \surname Fernandes \email jlaraf@terra.com.br \\
	\addr Jose Michel Kalaf Research Institute, Sao Paulo, Brazil
	\AND
	\firstname Christoph \surname Tillmanns \email tillmanns@diagnostikum-berlin.de \\
	\addr Diagnostikum Berlin, Berlin, Germany
	\AND
	\firstname Jens \surname Wetzl \email jens.wetzl@siemens-healthineers.com \\
	\addr Magnetic Resonance, Siemens Healthcare GmbH, Erlangen 91052, Germany
	\AND
	\firstname Andreas \surname Maier \email andreas.maier@fau.de \\
	\addr Pattern Recognition Lab, Department of Computer Science, Friedrich-Alexander University Erlangen-Nuremberg, Erlangen 91058, Germany 
}

\begin{document}

\maketitle

\vspace{-2em}
\begin{abstract}
		Static cardiac imaging such as late gadolinium enhancement, mapping, or 3-D coronary angiography require prior information, e.g., the phase during a cardiac cycle with least motion, called resting phase (RP). The purpose of this work is to propose a fully automated framework that allows the detection of the right coronary artery (RCA) RP within CINE series. The proposed prototype system consists of three main steps. First, the localization of the regions of interest (ROI) is performed. Second, the cropped ROI series are taken for tracking motions over all time points. Third, the output motion values are used to classify RPs. In this work, we focused on the detection of the area with the outer edge of the cross-section of the RCA as our target. The proposed framework was evaluated on 102 clinically acquired dataset at 1.5T and 3T. The automatically classified RPs were compared with the reference RPs annotated manually by a expert for testing the robustness and feasibility of the framework. The predicted RCA RPs showed high agreement with the experts annotated RPs with 92.7\% accuracy, 90.5\% sensitivity and 95.0\% specificity for the unseen study dataset. The mean absolute difference of the start and end RP was 13.6 $\pm$ 18.6 ms for the validation study dataset (n=102). In this work, automated RP detection has been introduced by the proposed framework and demonstrated feasibility, robustness, and applicability for static imaging acquisitions.
\end{abstract}


\section{Introduction}
    In cardiovascular magnetic resonance (CMR) imaging, static cardiac imaging techniques, such as late gadolinium enhancement (LGE) \citep{kellman2012cardiac,kellman2002phase, akccakaya2012accelerated,basha2017clinical}, mapping \citep{kellman2014t1,messroghli2017clinical,aherne2020cardiac}, or three-dimensional (3-D) whole heart coronary angiography \citep{munoz2020motion, greil20173d, cruz2017highly,univis91495003} are increasingly being performed to qualitatively and quantitatively assess the cardiac anatomy and function. It is important to acquire the data during the phase of the cardiac cycle with least motion, called a resting phase (RP), especially mid- or end-diastolic (ED) phases \citep{kramer2020standardized,isma2009optimal,kramer2013society}, or in patients with a fast heart rate during the end-systolic (ES) phase. \\
    In standardized CMR protocols \citep{kramer2020standardized, kramer2013society}, the guidelines recommend using the diastolic RP with a duration of less than 200 ms as the acquisition window for static cardiac imaging. In certain situations, e.g., high heart rate or patients with arrhythmias, especially in terms of mapping acquisition, the systolic RP is preferably chosen. As outlined in \citep{kim2001impact,seifarth2007optimal, hofman1998quantification},  electrocardiogram-based heuristics enable the ED phase selection based on a trigger time at 75\% of the RR interval for most patients, however it can be suboptimal due to the magnetohydrodynamic effect \citep{abi2007alterations}, and not generalizable, especially for patients with high or irregular heart rates. 
    For advanced applications such as high-resolution angiography, accurate determination of RP is necessary. As different structures in the heart rest at different times of the cardiac cycle, ideally a targeted RP for the anatomy of interest should be determined. For coronary angiography, for example, it is suggested to accurately determine the RP of the right coronary artery (RCA) \citep{kramer2020standardized, kramer2013society, shechter2005rest, johnson2004three, wang2001coronary}. \\
    The selection of the RPs is typically performed based on visual inspection on a CINE series acquired prior to the static imaging. In current clinical practice, an expert is required to select either the end-systolic, mid- or end-diastolic phase for acquisition. To tackle the complex and time-consuming manual task of RP selection for the static cardiac imaging, some previous studies have been introduced to perform the RP determination automatically.  
    In previously conducted studies \citep{wang2001coronary, stuber1999submillimeter}, a calibration scan-based approach using navigator echoes has been presented, however this approach requires significant user experience and interaction in order to accurately plan the navigator position. A threshold-based clustering algorithm was proposed to track low-motion periods \citep{suever2011time}. However to track the RCA area, manual tracking is required.
    Another approach \citep{jahnke2005new} was proposed using image based cross-correlation of CINE series for the automatic selection of RPs, proving to be advantageous in terms of image quality. An extension of the previous method \citep{ustun2007automated} calculates the myocardial displacement from the cross-correlation calculation. These methods, however, also require user interaction to position the region of interest (ROI) enclosing the heart.
    An alternative technique with automated RCA positioning \citep{sato2009approach} using template matching algorithm was proposed to automatically select RP based on image intensity differences. The intensity difference calculation as well as the template matching algorithm can be sensitive to artifacts.
    Further, a method attempts to determine the cardiac motion resolution-independently \citep{huang2014automatic} using intensity standard deviation calculation, and additionally proposed the motion extraction based on deformable model-based registration. Another idea was proposed in which the feasibility of motion correction algorithm for quantifying the rest periods of the coronary arteries is shown \citep{shechter2005rest}. However, the computed RP is based on the entire field-of-view and do not provide localized RPs for specific anatomies, and the detection is limited to two RPs due to the two local minima search. In a recent study \citep{asou2018automated}, an automated RP selection algorithm was introduced based on a motion area map generated from the high-speed component of the motion within a CINE series, however the high-speed component is not necessarily related to the anatomical structures. \\
    Deep Learning approaches have the potential to automate clinical workflows, and the state-of-the-art methods using convolutional neural networks (CNN) are currently used for image localization and segmentation \citep{krizhevsky2012imagenet, szegedy2015going,simonyan2014very, ronneberger2015u}. The CNN-based models are particularly used for learning the optimal spatial features from input data, especially images, thus performing a specific task such as localization or segmentation. Beside the power of CNN models for learning the spatial information, 3-D convolutional operators are used to learn spatial and temporal information for capturing features in a 3-D input data \citep{ji20123d, tran2015learning}. 
    Registration approaches are commonly used when it comes to motion analysis, object tracking, etc. \citep{dyke2019non,chefd2002flows, szeliski1996motion, spinei1998spatiotemporal, rueckert1999nonrigid}, by estimating a smooth correspondence function mapping between the coordinates from a reference image and those in a target image. These techniques can be used for calculating the motion of a target with the deformation fields within CINE series quantitatively.
    
    In this work, we propose a fully automated prototype system combining the advantages of the 3-D CNN and registration algorithms for detecting localized RPs of the RCA from 4-chamber view (4CH) CINE series. As the CINE series are time-resolved images, a 3-D CNN based model is trained to perform landmark detection over the cardiac phases. The proposed system combines the deep neural network for landmark detection and a registration algorithm. The motion within a localized anatomy is quantified in order to automatically classifying the systolic and diastolic RPs of the RCA. The duration of the RPs is quantified in both cases. To test the robustness and feasibility, the proposed framework was integrated into the scanner software and validated on patient data acquired on 1.5T and 3T scanners at multiple centers and different CINE sequences.

\section{Methods}

	\subsection{System Overview}
	\label{subsec:systemoverview}
		\begin{figure}[t]
    		\centering
    		\includegraphics[width=1\linewidth]{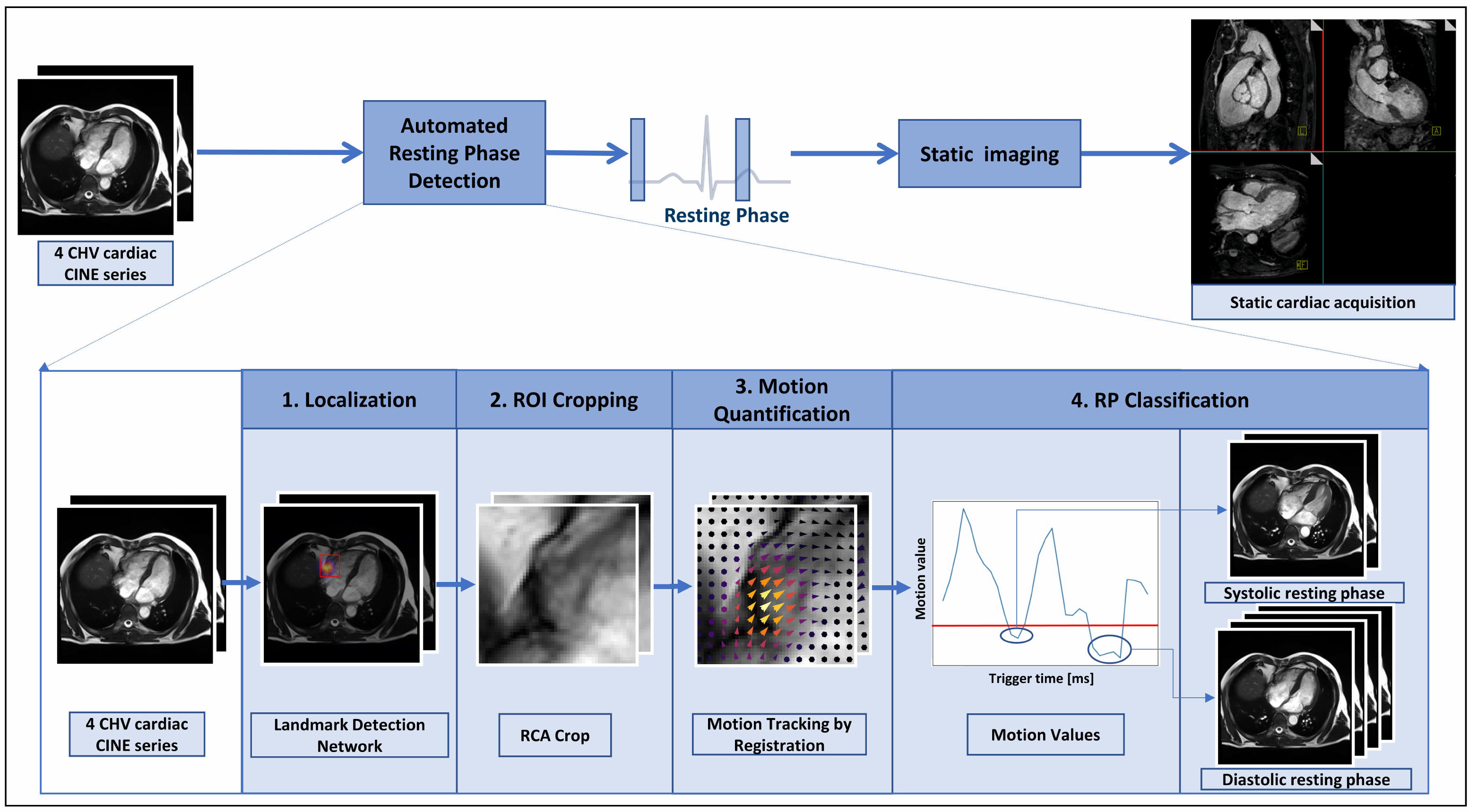}
    		\caption{Overview of the proposed system. Prior to a static imaging, a RP must be defined to prevent motion artifacts. The automated RP detection system consists of localization, ROI cropping, motion quantification and RP classification steps and provides the RPs of the target of interest within a 4CH CINE series.}
		  \label{fig:1}

	    \end{figure}
	    
		The proposed prototype system consists of three main steps that are executed consecutively (Figure \ref{fig:1}). The first step is to localize the ROI (see details in section \ref{subsec:localization}) from the input which is a 4CH CINE series. ROI can be chosen for any anatomical structures, that are of interest displayed in 4CH images, such as the RCA. The localization can be performed by neural networks trained for landmark detection tasks. In case of landmark detection, the output can be pixel coordinates that are used for cropping the image to the localized anatomy (see details in section \ref{subsec:cropping}). The cropped series containing the ROI are used for motion quantification (see details in section \ref{subsec:motionquantification}) by performing the elastic image registration \citep{chefd2002flows}. By taking the median of the magnitude of the deformation fields from consecutive frames, the motion values over time points are calculated. Finally, the RPs are classified by an absolute threshold, defined based on the correlations of the predicted RPs by varying the thresholds with the expert annotations (see details in section \ref{subsec:restingphaseclassification}). The frames which have lower motion values than the threshold value, are selected as RPs. The detected RPs are then used to plan the subsequent static image acquisition. 

	\subsection{Localization}
	\label{subsec:localization}
    	In this work, the landmark detection neural network is built to automatically detect the location of RCA in a 4CH time-resolved series. The densely connected neural network (3-D DenseNet) architecture is trained to regress the x- and y- coordinates of the RCA over time and it is described in detail here. 
        As a preprocessing step, the 4CH CINE series are interpolated to a fixed spatial and temporal size of $224\times224\times32$ to be independent from resolution. Further, the min-max pixel intensity normalization was applied to rescale the different intensity range in [0,1]. The 3-D DenseNet proposed in \citep{huang2017densely} was trained under supervised learning. The weights of the network were updated by using the Adam optimizer \citep{kingma2014adam} with $\lambda={10}^{-3}$ and the mean-squared-error loss function (MSE) as follows:
		\begin{align}
			MSE &= \frac{1}{N} \sum_{n=1}^{N} \biggl(\frac{1}{T} \norm{(\hat{y}_{t,n} - y_{t,n} }^2 \biggr)
			\label{eq:MSE}
		\end{align}
		\begin{table}[t] 
        \centering
        \caption{The extended 3-D DenseNet built for the RCA detection. Each Conv consists of the successively executed layers of 3-D batch normalization, rectified linear unit activation function and 3-D convolutions. For the RCA detection, the 3-D DenseNet with the total number of 122 convolutional layers are built as follows: $d=32$, $h,w=224$ and $c_{1,2,3,4} = {6,12,24,16}$.}
	    \resizebox{0.6\columnwidth}{!}{

        \begin{tabular}{ l||l||l }
         \toprule
        Layers & Output Size & 3-D DenseNet \\
         \midrule
        Convolution & $d \times \frac{h}{2}\times\frac{w}{2}$ & \vtop{\hbox{\strut  $3\times3\times3$ Conv, }\hbox{\strut $1\times2\times2$ stride} }  \\
        \midrule
        Pooling & $\frac{d}{2}\times\frac{h}{4}\times\frac{w}{4}$ & \vtop{\hbox{\strut  $3\times3\times3$ max-pool, }\hbox{\strut $2\times2\times2$ stride} }\\
        
        \midrule
        Dense block  & $\frac{d}{2}\times\frac{h}{4}\times\frac{w}{4}$ & 
                                                                        \(
                                                                        \begin{bmatrix} 
                                                                            1\times1\times1 \hspace{0.1cm} \text{Conv} \\ 
                                                                            3\times3\times3 \hspace{0.1cm} \text{Conv} 
                                                                        \end{bmatrix} \times  c_{1} \) \\

        \midrule 
        Transition block & $\frac{d}{4}\times\frac{h}{8}\times\frac{w}{8}$ & \vtop{\hbox{\strut $1\times1\times1$ Conv,} \hbox{\strut $2\times2\times2$ avg-pool,} \hbox{\strut $2\times2\times2$ stride}  }      \\
        \midrule
        Dense block & $\frac{d}{4}\times\frac{h}{8}\times\frac{w}{8}$ & 
                                                                          \(\begin{bmatrix} 
                                                                            1\times1\times1 \hspace{0.1cm} \text{Conv} \\ 
                                                                            3\times3\times3 \hspace{0.1cm} \text{Conv} 
                                                                        \end{bmatrix} \times  c_{2} \)\\ 
        \midrule
        Transition block & $\frac{d}{8}\times\frac{h}{16}\times\frac{w}{16}$ & \vtop{\hbox{\strut $1\times1\times1$ Conv,} \hbox{\strut $2\times2\times2$ avg-pool,} \hbox{\strut $2\times2\times2 $ stride}  } \\
        \midrule
        Dense block & $\frac{d}{8}\times\frac{h}{16}\times\frac{w}{16}$ & 
                                                                          \(\begin{bmatrix} 
                                                                            1\times1\times1 \hspace{0.1cm} \text{Conv} \\ 
                                                                            3\times3\times3 \hspace{0.1cm} \text{Conv} 
                                                                        \end{bmatrix} \times  c_{3} \)\\
        \midrule
        Transition block & $\frac{d}{16}\times\frac{h}{32}\times\frac{w}{32}$ & \vtop{\hbox{\strut $1\times1\times1$ Conv,} \hbox{\strut $2\times2\times2$ avg-pool,} \hbox{\strut $2\times2\times2 $ stride}  }  \\
        \midrule
        Dense block & $\frac{d}{16}\times\frac{h}{32}\times\frac{w}{32}$ & 
                                                                          \(\begin{bmatrix} 
                                                                            1\times1\times1 \hspace{0.1cm} \text{Conv} \\ 
                                                                            3\times3\times3 \hspace{0.1cm} \text{Conv} 
                                                                        \end{bmatrix} \times  c_{4} \)\\ 
        \midrule
        Classification block & $d\times4$ & \vtop{\hbox{\strut  3-D adaptive avg-pool, }\hbox{\strut $1\times1\times1$ Conv} } \newline \\
        \bottomrule

        \end{tabular}}
        \label{table:RCAarchitecture}

    \end{table}
        where $\hat{y}_{t,n}$ is the predicted pixel coordinates at the time $t$ from $n$ dataset, the $\hat{y}_{t,n}$ ground truth, $T$ the number of frames, and $N$ the number of the datasets. 
        The ground truth is generated in a semi-supervised manner, where the RCA pixel coordinates in the first frame are manually annotated and propagated to the next frames using the deformation fields describing the displacement between $\hat{y}_{t,n}$ and $\hat{y}_{t+1,n}$. The deformation field are generated by using the elastic image registration \citep{chefd2002flows}. Each frame was then corrected manually if the propagated coordinates were not accurate.
        The total number of convolutional layers (Conv) is 122, and before each Conv a 3-D batch normalization (BN) \citep{ioffe2015batch} and rectified linear unit (ReLU) activation functions \citep{nair2010rectified} are applied. After the initial 3-D Conv and max pooling operator, the feature maps were forwarded through 4 concatenated dense blocks (DB) and transition blocks (TB). The number of 4 concatenated DBs are set to 6, 12, 24, 16, in which after each DB, a TB is applied. In each DB, 2 Convs, each followed by 3-D BN and ReLU operators with the increase of the feature maps with 12 are applied. Each layer obtains additional inputs from all preceding layers and forwards the feature maps to all subsequent layers. In each TB, a Conv with BN and ReLU followed by 3-D average pooling operator is applied for spatial and temporal down-sampling. As the last step, a global average pooling and $1\times1\times1$ convolutional operator are used to regress the coordinates from the extracted features maps. The detailed architectural details can be found in the Table \ref{table:RCAarchitecture}.

    \subsection{Cropping}
	\label{subsec:cropping}
        From the output of the localization task, a ROI can be simply selected from the pixel coordinates. The detected pixel coordinates are transformed back to the original coordinate system, and then the cropping is performed. 
        Given the predicted pixel coordinates of RCA by 3-D DenseNet, the bounding box is defined by taking the minimum and maximum x- and y-coordinates of the points in the coordinate plane from a time-resolved series and calculating the average of these x and y-coordinates. The size of the bounding box is selected based on prior knowledge about the size of the anatomy, chosen as  $50\times50\ mm^2$. 

    \subsection{Motion Quantification}
    \label{subsec:motionquantification}
\begin{figure}[t]
    		\centering
    		\includegraphics[width=1\linewidth]{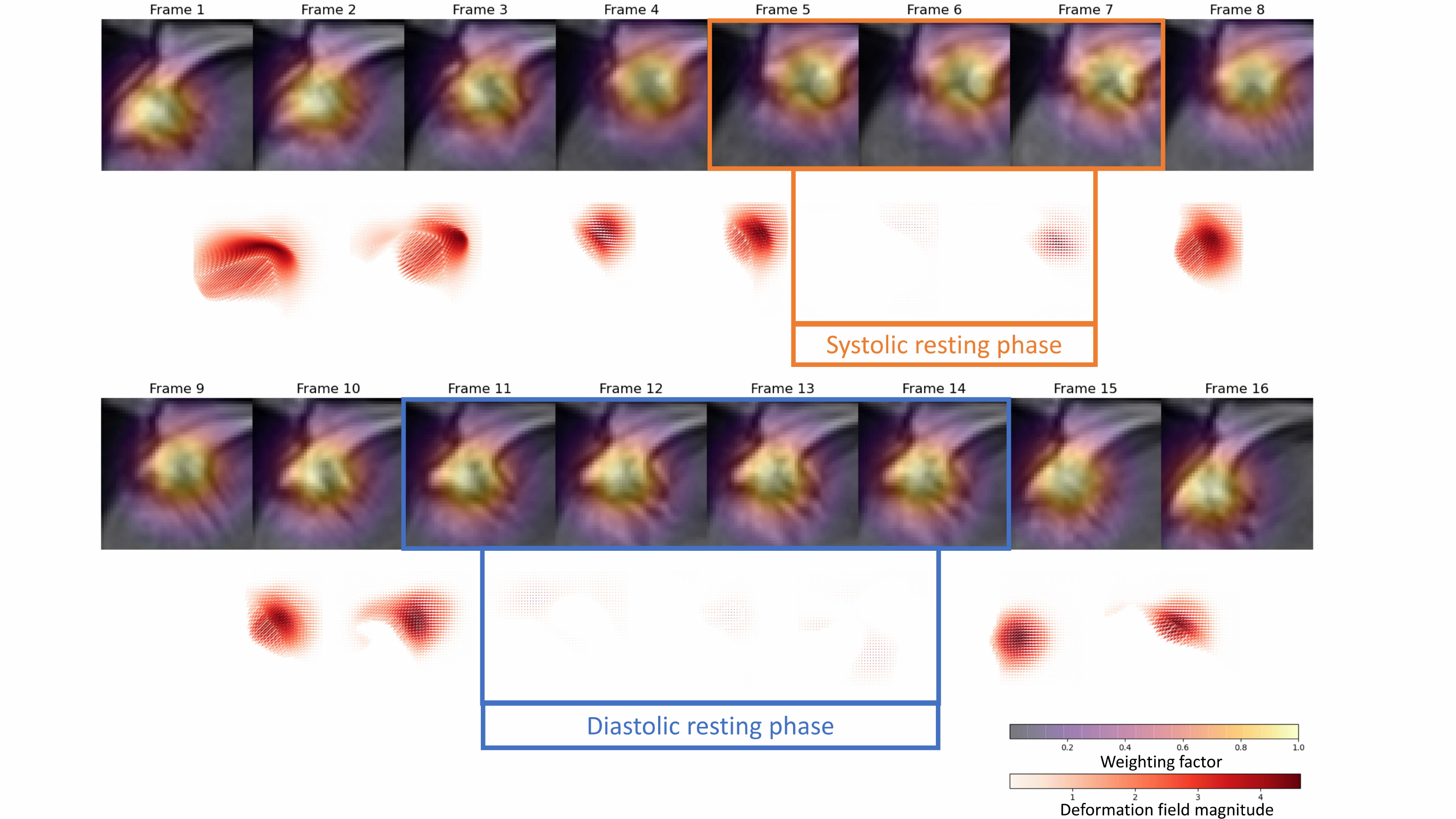}
    		\caption{An example of RCA ROI overlayed with weighted heatmaps, and below each two frames the magnitude of the weighted consecutive deformation fields are illustrated. The upper color bar corresponds to the weighted heatmaps, and the below one to the magnitude of the deformation vectors. The predicted systolic resting phase is marked in orange, and the predicted diastolic resting phase in blue. For the sake of simplicity, the last frame is not visualized.}
    	    \label{fig:2}
	    \end{figure}

        The motion values are quantitatively determined using elastic image registration \citep{chefd2002flows}. Consecutive frames of the CINE series, $s_t(x)$ and $s_{t+1}(x)$ for all timepoints $t$, are registered to obtain deformation fields $d_t(x)$ such that $s_{t+1}(d_t(x))$ minimizes the dissimilarity measure related to $s_t(x)$. The motion curve $m(t)$ describing the amount of RCA motion is then computed as the median of the weighted magnitudes of the deformation fields $\norm{d_t(x)}$  as follows:
		\begin{align}
			m(t) &= median\{G_t(x) \cdot \norm{d_t(x)}^2\}
			\label{eq:motion}
		\end{align}
        where $G_t(x)$ is a Gaussian weighting function centered at the midpoint of the detected location of the RCA between $\hat{y}_t$ and ${\hat{y}}_{t+1}$  at the time point t:

		\begin{align}
			G_t(x) &= \exp\biggl(-\frac{\norm{x-p_t}^2}{\sigma^2}\biggr)
			\label{eq:gaussian}
		\end{align}
        while $p_t$ denotes the midpoint of the detected RCA position. This Gaussian weighting ensures that the motion curve represents mainly the motion of the RCA, while still being robust to slight imprecisions of the localization results. The standard deviation was empirically chosen as $\sigma=12$. Figure \ref{fig:2} shows the Gaussian weighting functions overlaid on the anatomical images at each time point, as well as the weighted deformation fields corresponding to subsequent image pairs.
        The quantification of motion can be considered in different ways, and the detailed analysis of obtaining the RCA motion values can be found in section \ref{subsec:results_motionquantifcation}.
    \subsection{Resting Phase Classification}
    \label{subsec:restingphaseclassification}
        After the motion curve is quantified, a classification is required to obtain the RP. This classification is performed within a time interval $T_{valid}=[\alpha, \omega]$, i.e. the first $\alpha$ and the last $\omega$ ms of a cardiac cycle are excluded. $\alpha$ accounts for the time needed for preparation pulses before the data acquisition window can start. $\omega$ is a safety margin at the end of the cardiac cycle to account for heart rate variability, i.e. the detected resting phase should not be too late in the cardiac cycle in case subsequent cardiac cycles during the 3-D acquisition are shorter, which can lead to unstable measurements \citep{kim2001impact, seifarth2007optimal}.
        Based on the ground truth annotation of RPs, the $\alpha$, and $\omega$ are empirically chosen to be 80 ms. 
        The RPs can be determined with an absolute threshold from the motion values. The frames with motion values lower than the absolute threshold value are assigned as RPs which can be described as follows:

		\[
		    RP(t) = 
		    \begin{cases}
		        1, & m(t) < \tau, t\in T_{valid} \\
		        0, & m(t) \geq \tau, t\in T_{valid}
		    \end{cases}
		\]
        where $\tau$ is the absolute threshold value, and $m(t)$ is the obtained motion value at trigger time $t$.
        The threshold value is obtained based on sensitivity (true positive rate) and specificity (true negative rate) analysis from manual annotations. The optimal absolute threshold is chosen as the one achieved with best balanced accuracy. 
    \subsection{Data}
    \label{subsec:data}
        \subsubsection{Training and Validation Dataset for the RCA Detection Network}
        Data used for training and evaluating the RCA detection network was acquired on 1.5T and 3T clinical MRI scanners (MAGNETOM Aera, Avanto, Prisma, Skyra, Trio TIM; Siemens Healthcare, Erlangen, Germany) at multiple centers (n=1000). The dataset was split into 70\% training, 15\% validation and 15\% testing set for the RCA detection.
        Details about the datasets used for training, validating, testing the RCA detection network, and evaluating the classified resting phases are shown in Table \ref{table:dataset}. A medical expert with more than 10 years of cardiac MRI experience manually annotated the RPs on 76 datasets from the testing set for the analysis of the system.
        \subsubsection{Study Dataset}
        Data used for evaluating the proposed system was acquired on 1.5T and 3T clinical MRI scanners (MAGNETOM Aera, Avanto fit, Skyra, Skyra fit, Sola, Vida; Siemens Healthcare, Erlangen, Germany) at multiple centers (n=102). The proposed system was integrated into the scanner software and tested online. The dataset from the study was not used for training, testing the RCA network or the threshold analysis, and was not mixed with the RCA Detection Network dataset. The test datasets consist of non-pre-selected patient data with a minimum heart rate of 35 and a maximum heart rate of 97 who are clinically referred for coronary assessment. None of the patients have not received beta-blockers or a target heart rate was determined. Details about the study datasets are listed in Table \ref{table:dataset}. 
        \begin{table}[t] 
		    \centering
		    \caption{Statistics about the data population and acquisition used for training and testing the RCA detection network, and of the additional unseen study dataset.}
		    \resizebox{0.6\columnwidth}{!}{
            \begin{tabular}{ p{3.5cm}||p{3.5cm} |p{3.5cm}  }
                \toprule
                & RCA Detection\newline Network Dataset \newline (Resting Phase \newline annotation) & Study Dataset  \\
                \midrule
                Number of \newline patients & 1000 \newline (76) & 102 \\
                \midrule
                Age & 55.0 $\pm$ 19.0 \newline (59.0 $\pm$ 17.2) & 39.3 $\pm$ 10.9\\
                \midrule
                Gender & 64 \% male \newline (68\% male) & 69\% male\\
                \midrule
                Heart Rate \newline [bpm] & 64.7 $\pm$ 11.6 \newline (68.0 $\pm$ 13.2) & 70.1 $\pm$ 12.3\\
                \toprule
                
                Field Strength &  $25\,\%\,1.5\,T$, $75\,\%\,3\,T$ \newline ($46\,\%\,1.5\,T$, $54\,\%\,3\,T$) & $22\,\%\,1.5\,T$, $78\,\%\,3\,T$\\
                \midrule
                Spatial Resolution \newline $[mm^2]$ & 1.4 $\pm$ 0.1 \newline (1.5 $\pm$ 0.2) & 1.7 $\pm$ 0.1\\
                \midrule
                Temporal Resolution \newline $[ms]$ & 33.8 $\pm$ 9.9 \newline (35.8 $\pm$ 6.8) & 37.3 $\pm$ 7.8\\
                \midrule
                FOV \newline[mm x mm] & 311.3 $\pm$ 30.6 \newline 344.4 $\pm$ 26.9 \newline (312.2 $\pm$ 31.7 \newline 357.4 $\pm$ 23.6) & 283.5 $\pm$ 9.2 \newline 345.9 $\pm$ 8.9\\
                \midrule
                Number of Frames & 25.8 $\pm$ 2.3 \newline (25.6 $\pm$ 1.6) & 26.0 $\pm$ 2.0\\
                
                \bottomrule
            \end{tabular}}
        \label{table:dataset}
        \end{table}
    
    \subsection{Experiments}
    \label{subsec:experiments}
        \subsubsection{Localization}
        The RCA detection is validated by calculating the mean and standard deviation of the Euclidean distance between the predicted pixel coordinates and the ground truth pixel coordinates.
		\begin{align}
		    \text{Distance Error} =\frac{1}{N} \sum_{n=1}^{N} \biggl( \frac{1}{T} \sum_{t=1}^{T} \norm{\hat{p}_{t,n} - p_{t,n}}^2 \biggr)
			\label{eq:distance_error}
		\end{align}
        where $N$ denotes the number of annotated test dataset, $T$ the number of time frames in each CINE series and $\hat{p}$ the predicted and $p$ the ground truth RCA position. The robustness and performance of the network was qualitatively validated on 12 oblique and diverse oriented 4CH CINE and 9 clinically acquired unseen study dataset with different field strength scanners. Further, the network was evaluated by a box plot showing the performance of the prediction at each frame.  
        
        \subsubsection{Motion Quantification}
        In order to find the best approach to quantify motion values, several approaches were evaluated on the annotated datasets for quantifying RCA motion from a cropped CINE series. The first approach is to quantify motion based on the distance between detected pixel coordinate over each adjacent time point as follows:
		\begin{align}
		    m_{\text{dist}(t)} &=\norm{p_t - p_{t+1}}
			\label{eq:m_dist}
		\end{align}
        The second is to aggregate the magnitudes of the deformation fields within the ROI without the Gaussian weighting using percentile or mean: 
		\begin{align}
		    m_{\text{pct}}(t) &= \eta_n \{\norm{d_t(x)} |\, x \in \text{ROI, or\}}
			\label{eq:m_pct}
		\end{align}
		\begin{align}
		    m_{\text{mean}}(t) &= mean\{\norm{d_t(x)} |\, x \in \text{ROI\}}
			\label{eq:m_mean}
		\end{align}
        where $\eta_n$ is the $n^{th}$ percentile. \\ Our last proposed approach is to aggregate the weighted deformation field magnitudes to calculate the motion values as described in the following:
		\begin{align}
		    m_{\text{wpct}}(t) &= \eta_n \{G_t(\vec{x})\cdot \norm{d_t(x)} |\, x \in \text{ROI, or\}}
			\label{eq:m_wpct}
		\end{align}
		\begin{align}
		    m_{\text{wmean}}(t) &= mean\{G_t(x)\cdot \norm{d_t(x)} |\, x \in \text{ROI\},}
			\label{eq:m_wmean}
		\end{align}
        where $\norm{d_{t}(x)}$ is the magnitudes of the deformation fields.
        The percentile analysis is performed to quantify motion from the deformation fields, which is based on calculating the balanced accuracy, sensitivity and specificity by varying $n$ from \nth{10} to \nth{100} by \nth{10} percentile steps. Further, the mean value is calculated as well, and compared with the percentile analysis.
        In addition, the motion values of the above mentioned 9 clinically acquired dataset are extracted and qualitative validated.
        
        \subsubsection{Threahold Analysis}
        The analysis for finding the optimal threshold value to determine the RPs from the quantified motion value is performed by the binary classification task with varying the threshold $\tau$ from 0.01 to 1 by 0.01 steps by calculating the sensitivity and specificity. The analysis is performed separately for 1.5T and 3T and using the annotated datasets. The performance with the selected threshold $\tau$ is analyzed based on area under curve (AUC) from the receiver operating characteristic (ROC) curve \citep{swets1979roc}, and confusion matrices are evaluated on testing and study datasets.
        For all different approaches of motion quantification, the threshold analysis is performed such that each approach can be fairly compared. Based on thresholding, the accuracy of classified RPs is evaluated as described in section 2.5.
        
        \subsubsection{RP Classification}
        To evaluate the performance of RP classification, the mean absolute error (MAE) and the standard deviation between the system predicted $\widehat{RP}$ and annotated start and end time points of systolic and diastolic RP  are calculated as follows:
		\begin{align*}
		    MAE_{\lambda,\mu_{\text{type}}} &= \frac{1}{N} \sum_{n=1}^{N} \bigg|\widehat{RP}_{\lambda,\mu_{N,\text{type}}} - RP_{\lambda,\mu_{N,\text{type}}} \bigg|, \\
		    &\text{where}\, \lambda \in \text{start, end of RP}, \mu \in \text{window, frame}
			\label{eq:MAE}
		\end{align*}
        The number of RP annotated dataset is denoted as $N$, and the type can be the classified systolic (sys) or diastolic (dia) RP. The performance was validated by two different measures, firstly the difference of the time window and secondly, the number of images between the predicted and ground truth annotation. The time window specifies the accuracy in milliseconds, whereas the frame in number of images.
        The validation was performed on the testing datasets in which the RPs are manually annotated by a medical expert used for the threshold analysis. The performance of the system was analyzed based on Bland-Altmann analysis. The RPs was not counted when it was very short (\textless 30 ms, n=10), i.e., not resolvable by the temporal resolution of the acquisition.
        Additionally, the results of the system validation were presented by sensitivity, specificity and accuracy. To overcome the imbalanced classes (RP, no RP), the balanced accuracy is calculated based on true positive rate, and false true negative rate as follows: Accuracy = $\frac{(TPR+TNR)}{2}$, where TPR is true positive rate, and TNR true negative rate. Further, the ranges of each annotated RP type and predicted RP type were compared. \\

        To evaluate the robustness of the proposed system, different CINE sequences (Cartesian segmented, Cartesian segmented with small field-of-view, Cartesian segmented Compressed Sensing Prototype (CS), Cartesian Real-time CS, Radial real-time) were acquired and the predicted RPs of each sequence were compared with each of the expert annotation. Further, the computation time of the proposed system was measured at the beginning and the end of the proposed system.

\section{Results}
\label{sec:results}
    \subsection{Localization}
    \label{subsec:results_localization}
  
\begin{figure}[t]
    		\centering
    		\includegraphics[width=1\linewidth]{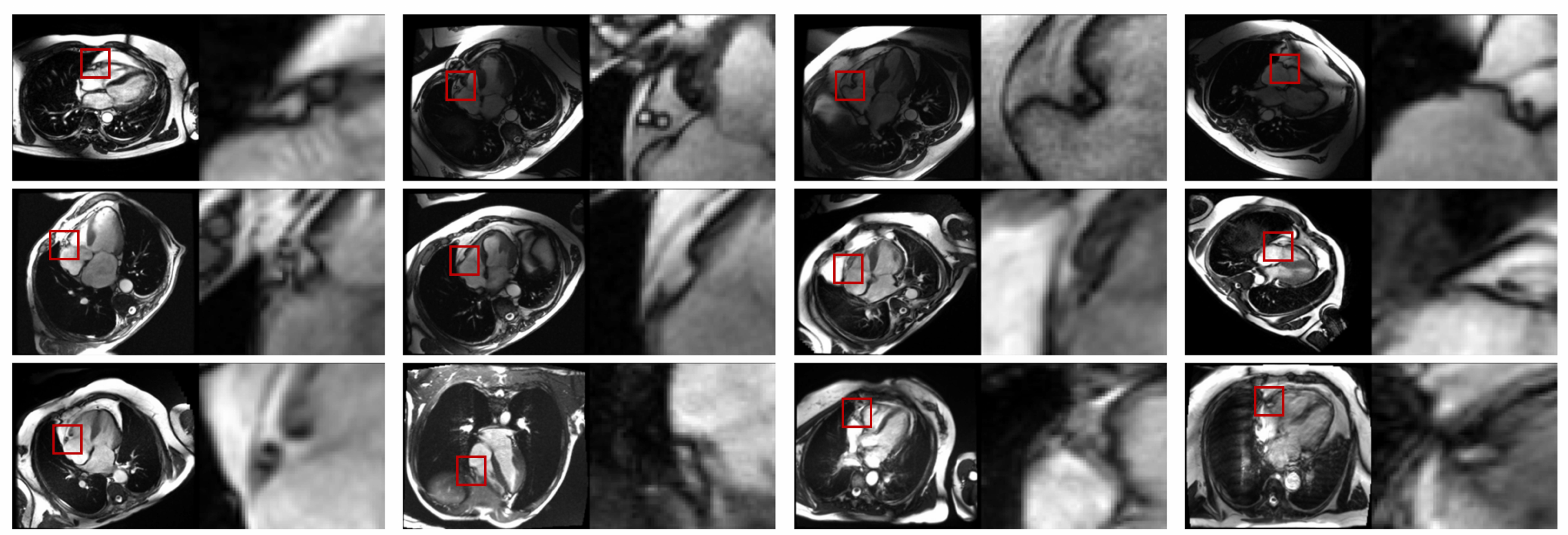}
    		\caption{The robustness of the RCA detection network is shown in a subset of the unseen dataset. For each case, the 4CH input is shown with a bounding box in red and the cropped image next to it for a total of 12 cases.}
    		\label{fig:oblique}
	    \end{figure}

\begin{figure}[t]
    		\centering
    		\includegraphics[width=0.8\linewidth]{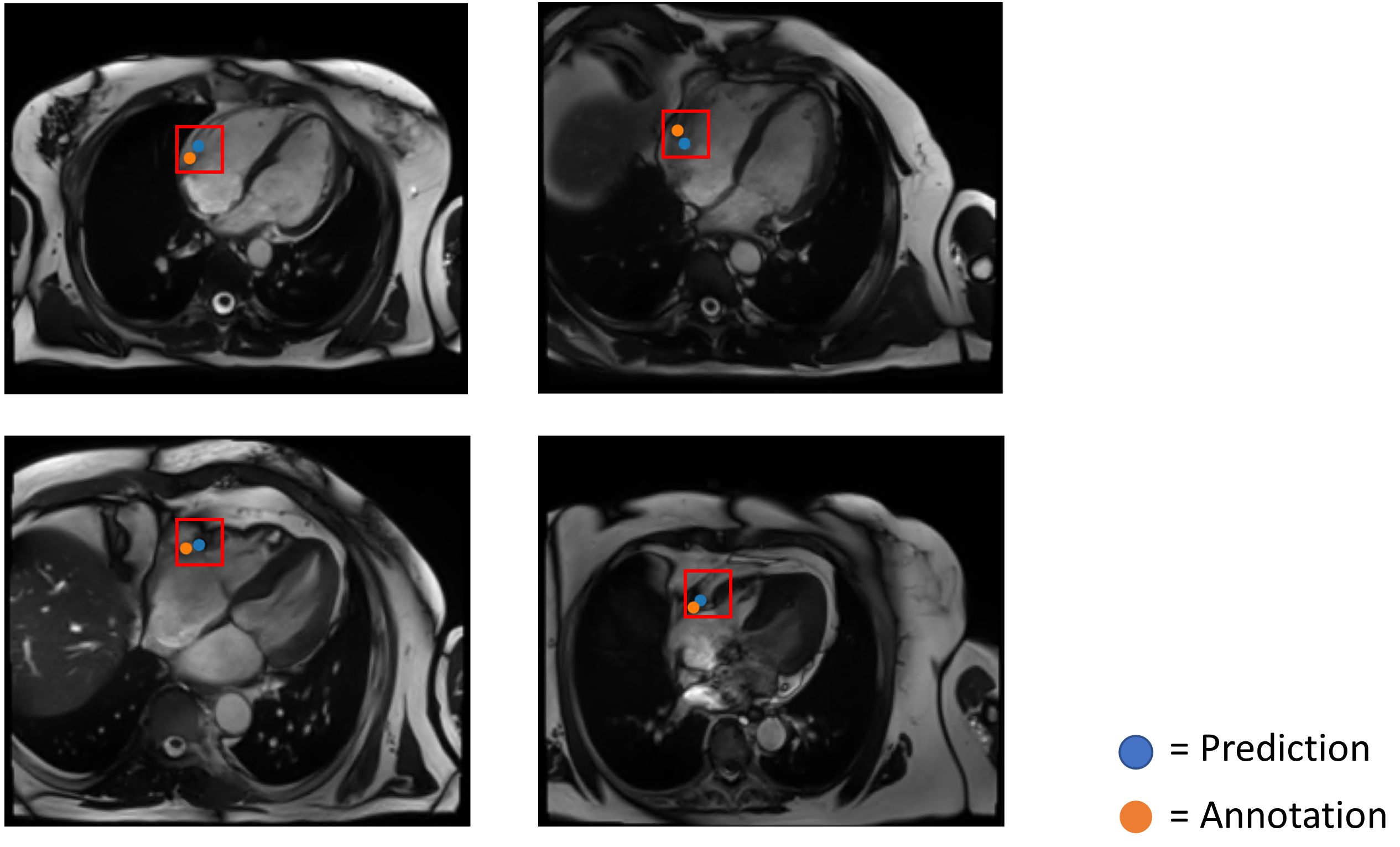}
    		\caption{The worst cases on the test dataset of the RCA detection network are visualized (n=4). The blue dot represents the predicted landmark position, the orange dot, the expert's annotation and the red box, the ROI defined based on the predicted landmark positions, respectively. Despite incorrect detection, the annotation landmark is still invariably included in all ROI windows.}
    		\label{fig:worst_cases}
	    \end{figure}

\begin{figure}[t]
    		\centering
    		\includegraphics[width=1\linewidth]{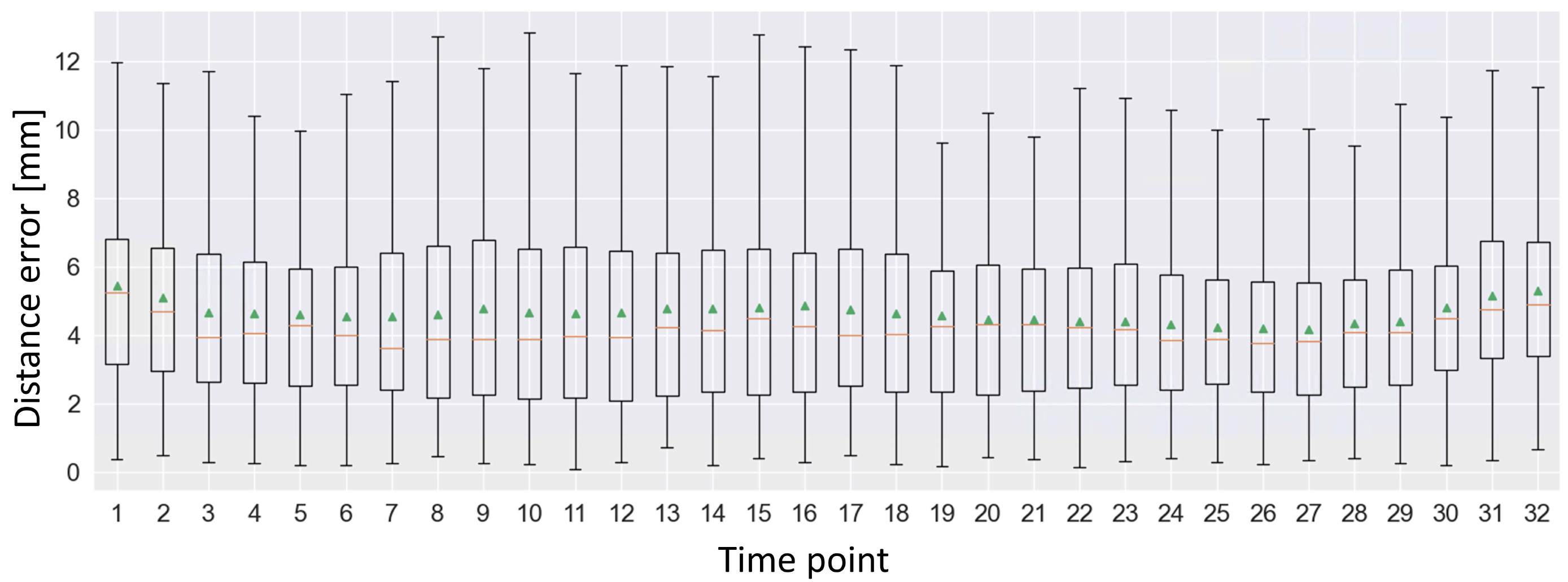}
    		\caption{A box plot showing the performance of the 122-layer 3-D DenseNet in each time point is illustrated. Orange line represents the median value and the green triangle, the mean value.}
    		\label{fig:boxplot}
	    \end{figure}

\begin{figure}[tp]
            \centering
    		\begin{subfigure}{0.9\textwidth}
    		\includegraphics[width=\textwidth]{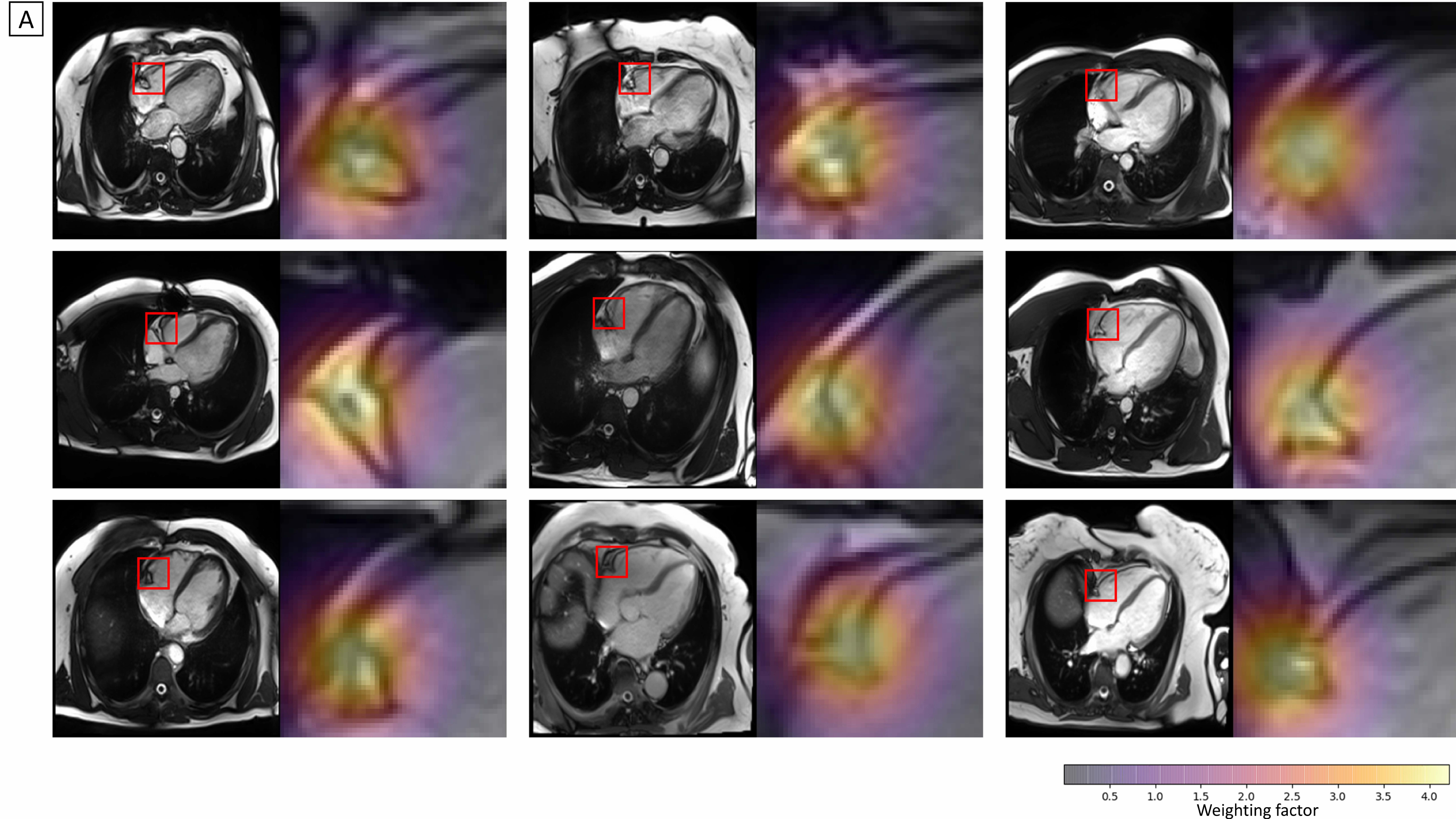}
            
            \end{subfigure}
            \vfill
    		\begin{subfigure}{0.9\textwidth}
    		\includegraphics[width=\textwidth]{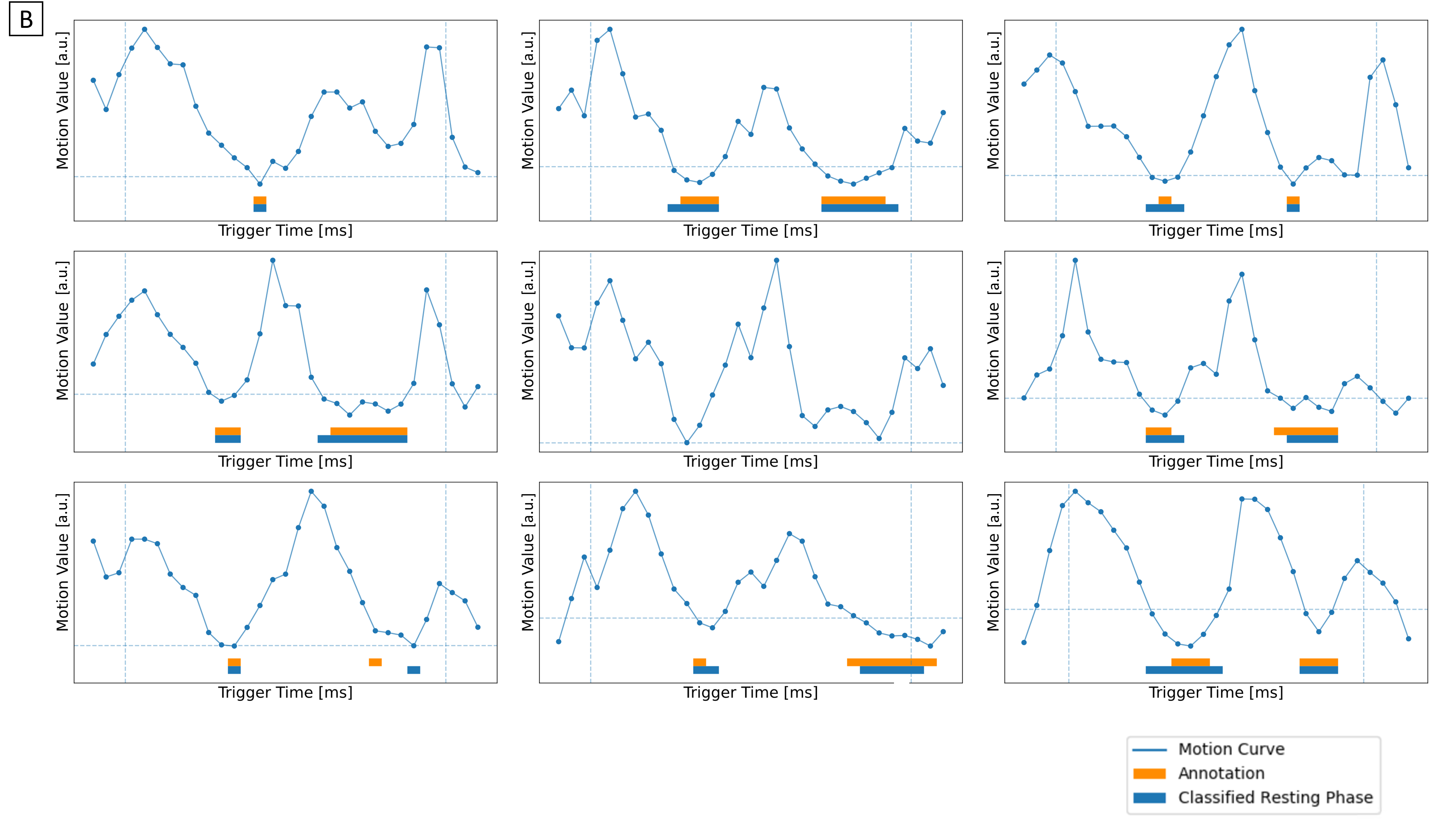}
            
            \end{subfigure}
    		\caption{Overview of the first frame of 9 different RCA series highlighted with a generated heatmap, which center point is taken from the predicted coordinate by the 3-D DenseNet trained for detecting the RCA pixel coordinate (A). The quantified motion value and classified RPs corresponding to each dataset are shown below (B). The vertical dashed line represents the window of interest as described in     \ref{subsec:restingphaseclassification}. The horizontal dashed line represents the selected threshold value. The lower the motion value, the less motion exists in the frame.}
    		\label{fig:3}
	    \end{figure}

		The mean and standard deviation error between the prediction and ground truth of the fully convolutional 3-D DenseNet with 122 layers was 4.6 ± 1.8 mm. The box plot in Figure \ref{fig:boxplot}, shows the Distance Error in mm between the $\hat{p}$ and the $p$ in each frame. 
		Robustness results for unseen datasets are presented in two ways, first in Figure \ref{fig:oblique} which shows the performance on oblique and diverse oriented cases (n=12) and second in Figure \ref{fig:worst_cases}, which visualized the worst cases.
		The quantitative localization results of a part of the study dataset (n=9) acquired with breath-hold and free-breathing CINE sequences are shown in Figure \ref{fig:3} above. The first frame of each CINE series and the corresponding RCA cropped series are shown. Each case was visualized with the first frame of the corresponding CINE series marked with a red box showing the position of ROI defined based on the network prediction and beside it with the cropped series enclosing the area with the outer edge of the cross-section of the RCA overlayed with the generated heatmap.   
	    
    \subsection{Motion Quantification}
    \label{subsec:results_motionquantifcation}
    \begin{table}[t] 
        \centering
        \caption{The analysis of motion quantification. Two approaches by using the Gaussian weighting of the magnitude of the deformation fields, or without any weighting have been compared for varying the percentile and taking the mean value.}
        \resizebox{0.7\columnwidth}{!}
        {
           \begin{tabular}{ c|c||c|c|c|c}
             \toprule
             Metric & Percentile & Accuracy & Sensitivity & Specificity & Threshold [a.u.]\\
             \midrule
             $m_{\text{dist}}$ & - & 71.4 & 50.5 & 92.3 & 0.50 \\
             \midrule
             $m_{\text{pct}}$ & \nth{10} & 75.1 & 55.5 & 94.6 & 0.08 \\
             $m_{\text{pct}}$ & \nth{20} & 81.5 & 67.3 & 95.8 & 0.13 \\
             $m_{\text{pct}}$ & \nth{30} & 84.1 & 72.2 & 96.0 & 0.18 \\
             $m_{\text{pct}}$ & \nth{40} & 86.9 & 77.6 & 96.2 & 0.23 \\
             $m_{\text{pct}}$ & \nth{50} & 87.2 & 78.1 & 96.4 & 0.28 \\
             $m_{\text{pct}}$ & \nth{60} & 87.7 & 80.1 & 95.3 & 0.34 \\
             $m_{\text{pct}}$ & \nth{70} & 86.5 & 76.2 & 96.7 & 0.38 \\
             $m_{\text{pct}}$ & \nth{80} & 83.2 & 69.0 & 97.5 & 0.42 \\
             $m_{\text{pct}}$ & \nth{90} & 81.7 & 65.8 & 97.7 & 0.49 \\
             $m_{\text{pct}}$ & \nth{100} & 78.3 & 61.6 & 95.1 & 0.68 \\
            $m_{\text{mean}}$ & - & 89.3 & 84.7 & 93.9 & 0.34 \\
             \midrule
            $m_{\text{wpct}}$ & \nth{10} & 62.4 & 29.5 & 95.2 & 0.01 \\ 
            $m_{\text{wpct}}$ & \nth{20} & 72.8 & 51.9 & 93.6 & 0.03 \\
            $m_{\text{wpct}}$ & \nth{30} & 82.7 & 74.2 & 91.2 & 0.07 \\
            $m_{\text{wpct}}$ & \nth{40} & 87.4 & 80.8 & 94.0 & 0.12 \\
            
            \boldmath$m_{\text{\bfseries wpct}}$ & \bfseries\nth{50} & \bfseries90.1 & \bfseries85.4 & \bfseries94.8 & \bfseries0.20 \\
            $m_{\text{wpct}}$ & \nth{60} & 88.0 & 78.9 & 97.2 & 0.28 \\
            $m_{\text{wpct}}$ & \nth{70} & 85.8 & 74.4 & 97.2 & 0.42 \\
            $m_{\text{wpct}}$ & \nth{80} & 85.1 & 73.4 & 97.0 & 0.62 \\
            $m_{\text{wpct}}$ & \nth{90} & 83.6 & 71.2 & 96.0 & 0.98\\
            $m_{\text{wpct}}$ & \nth{100} & 66.4 & 33.2 & 99.5 & 1.0 \\
            
            $m_{\text{wmean}}$ & - & 86.2 & 76.6 & 95.8 & 0.36\\
            \bottomrule
            \end{tabular}
        }
        \label{appendix:table2}
        \end{table}
        The motion values obtained by calculating the distance between the predicted pixel coordinates in each adjacent time points achieved 61.1\% accuracy for 1.5T, and 52.8\% for 3T. As the motion quantification analysis in the Table \ref{appendix:table2} shows, the \nth{50} percentile/median of Gaussian weighting achieved 90.1\% accuracy, whereas the accuracy was 87.2\% without weighting the deformation field. The median performed 91.0\% accuracy for 1.5T and 88.9\% for 3T. The best accuracy achieved by taking the mean metric was 89.3\%, without Gaussian weighting. The motion values quantified based on the median approach with Gaussian weighting in 9 clinically acquired dataset are shown in Figure \ref{fig:3} bottom.
    \subsection{Threshold Analysis}
    \label{subsec:results_threshold_analysis}
\begin{figure}[tp]
    		\centering
    		\includegraphics[width=1\linewidth]{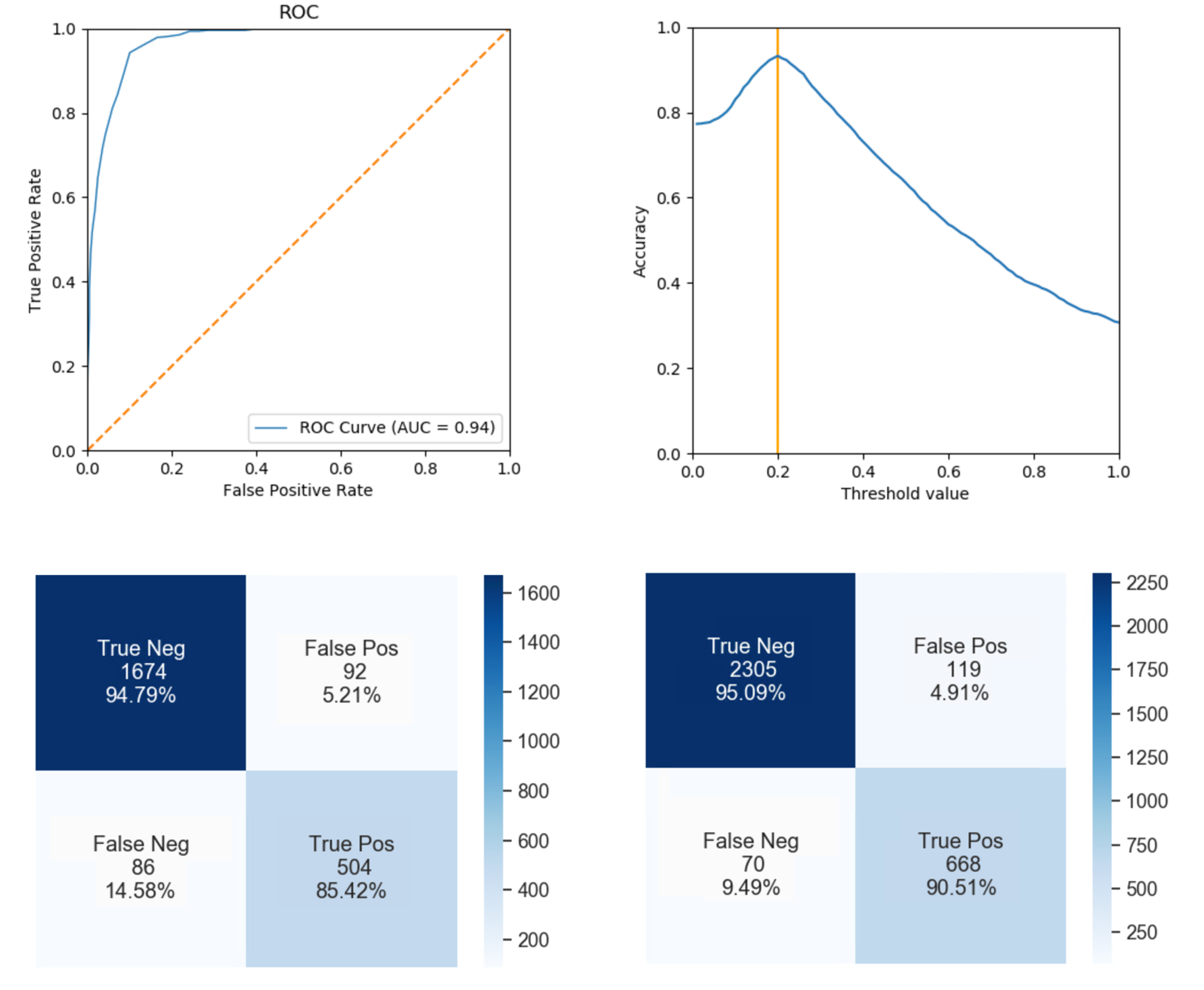}
    		\caption{Upper Left: the ROC curve from the threshold analysis is shown. Upper Right: the accuracy plot over the threshold values is shown. The best performed threshold value, 0.2 is marked in orange vertical line. Below Left:  the confusion matrix showing the performance of binary classification task by taking the best threshold value on testing datasets. Below Right: the confusion matrix showing the performance by taking the best threshold value on study datasets. In each measure, the counts and rate in percentage are listed.}
    		\label{fig:4}
	    \end{figure}

        For each percentile and mean analysis, the threshold $\tau$ selected based on binary classification task is listed in the right column in the Table \ref{appendix:table2}. From the motion values obtained by taking the median, the selected $\tau$ was 0.2. The resulting ROC curve is plotted in Figure \ref{fig:4}, and the accuracy over each threshold step is plotted in Figure \ref{fig:4} in the top row, in which the threshold $\tau$ is marked by an orange vertical line. On the below row in Figure \ref{fig:4} shows the confusion matrices of each annotated datasets. On the above one, the performance of the threshold on the testing dataset is shown, while on the below one the performance of the threshold on the study dataset is displayed.
    \subsection{RP Classification}
    \label{subsec:results_rpclassification}
        \begin{table}[h] 
    		\centering
    		
    		\caption{The results of the system on the unseen study datasets are listed. In the first row, the difference between the start and end RP and expert annotations are shown. In the second row, the difference of start and end systolic and diastolic RPs are shown in frame.}
            \resizebox{\columnwidth}{!}
            {
    	    \begin{tabular}{ p{2.5cm}|p{1.5cm}|p{1.8cm}||p{2.5cm}|p{2.5cm}|p{2.5cm}|p{2.5cm} }
                \toprule
                Magnetic Field & Number datasets & Threshold & Start  Systolic \newline RP  $[ms]$ & End  Systolic \newline RP  $[ms]$ & Start  Diastolic \newline RP  $[ms]$ & End  Diastolic\newline RP  $[ms]$\\
                \midrule
                $1.5\,T$ & N=22 & 0.20 & 
                14.4 $\pm$ 19.8 &
                15.0 $\pm$ 16.9 & 
                5.7 $\pm$ 13.1 & 
                12.2 $\pm$ 16.1 \\
                \midrule
                $3\,T$ &N=80 & 0.20 & 
                19.7 $\pm$ 22.4 &
                12.2 $\pm$ 16.6 & 
                15.8 $\pm$ 21.3 & 
                8.9 $\pm$ 15.2 \\
                \midrule
                $1.5\,T$ \& $3\,T$& N=102 & 0.20 & 
                18.7 $\pm$ 22.1 &
                12.7 $\pm$ 16.7 & 
                13.2 $\pm$ 20.0 & 
                9.7 $\pm$ 15.5 \\
                \bottomrule
            \end{tabular}
            }
            
            \vspace*{0.3 cm}
            \resizebox{\columnwidth}{!}
            {
            \begin{tabular}{ p{2.5cm}|p{1.5cm}|p{1.8cm}||p{2.5cm}|p{2.5cm}|p{2.5cm}|p{2.5cm} }
             \toprule
             Magnetic Field & Number datasets & Threshold & Start  Systolic \newline RP  $[Frame]$ & End Systolic\newline RP  $[Frame]$ & Start Diastolic\newline   RP $[Frame]$ & End Diastolic \newline RP  $[Frame]$\\
             \midrule
             $1.5\,T$ & N=22 & 0.20 & 
             0.46 $\pm$ 0.63 &
             0.54 $\pm$ 0.63 & 
             0.16 $\pm$ 0.36 & 
             0.37 $\pm$ 0.48\\
              \midrule
             $3\,T$ &N=80 & 0.20 & 
             0.70 $\pm$ 0.82 &
             0.43 $\pm$ 0.59 & 
             0.49 $\pm$ 0.63 & 
             0.28 $\pm$ 0.49\\
             
             \midrule
             $1.5\,T$ \& $3\,T$& N=102 & 0.20& 
             0.66 $\pm$ 0.80 &
             0.45 $\pm$ 0.60 & 
             0.40 $\pm$ 0.59 & 
             0.31 $\pm$ 0.49 \\
             \bottomrule
            \end{tabular}
            }
            
        \label{table:results_inRP_inFrame}
	    \end{table}
        \begin{table}[h] 
    		\centering
    		
    		\caption{Accuracy, sensitive and specificity are listed with the optimally defined threshold.}
            \resizebox{\columnwidth}{!}
            {
            \begin{tabular}{ p{2.5cm}|p{1.5cm}|p{1.8cm}||p{2cm}|p{2cm}|p{2cm} }
            \toprule
            Magnetic Field &Number datasets & Threshold & Accuracy & Sensitivity & Specificity \\
            \midrule
            $1.5\,T$ & N=22 & 0.20 & 
            93.4 &
            90.1 &
            96.8 \\
            
            \midrule
            
            $3\,T$ &N=80 & 0.20 & 
            92.6 &
            90.7 & 
            94.5 \\
            
            \midrule
            $1.5\,T$ \& $3\,T$ &N=102 & 0.20 &
            92.7 &
            90.5 &
            95.0 \\
            
            \bottomrule
            \end{tabular}
            }
        \label{table:results_acc}
	    \end{table}
	    
\begin{figure}[tp]
    		\centering
    		\includegraphics[width=1\linewidth]{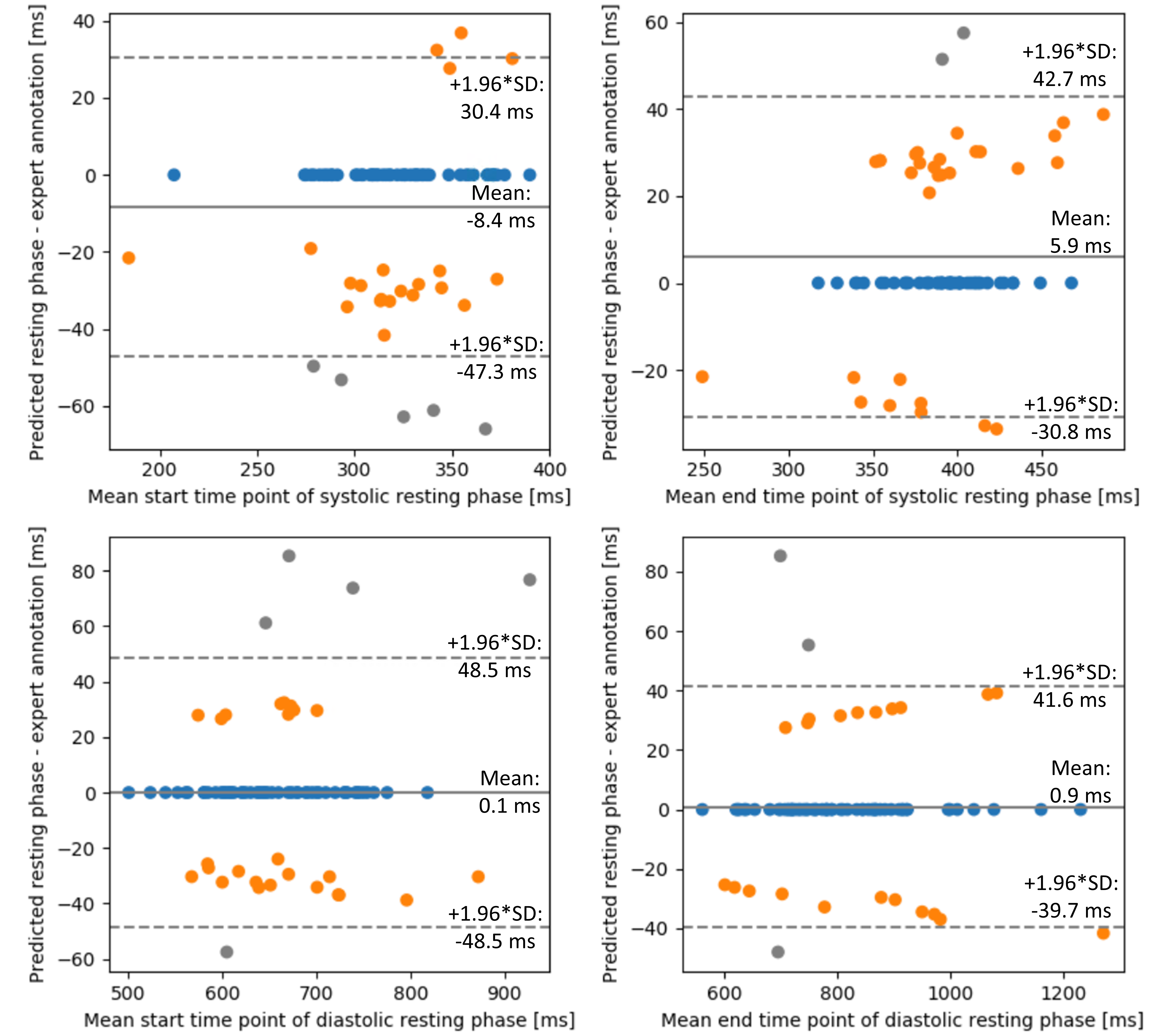}
    		\caption{Top: the difference between the predicted and expert annotations of start and end systolic RP are shown in Bland-Altmann plot. Bottom: the difference between the predicted and expert annotations of start and end diastolic RP are shown. The blue dots represent the exact match between the predicted and annotation, and the orange dots show when there is one frame difference. The gray dots represent when the difference is more than 2 frames.}
    		\label{fig:5}
	    \end{figure}

\begin{figure}[t]
    		\centering
    		\includegraphics[width=1\linewidth]{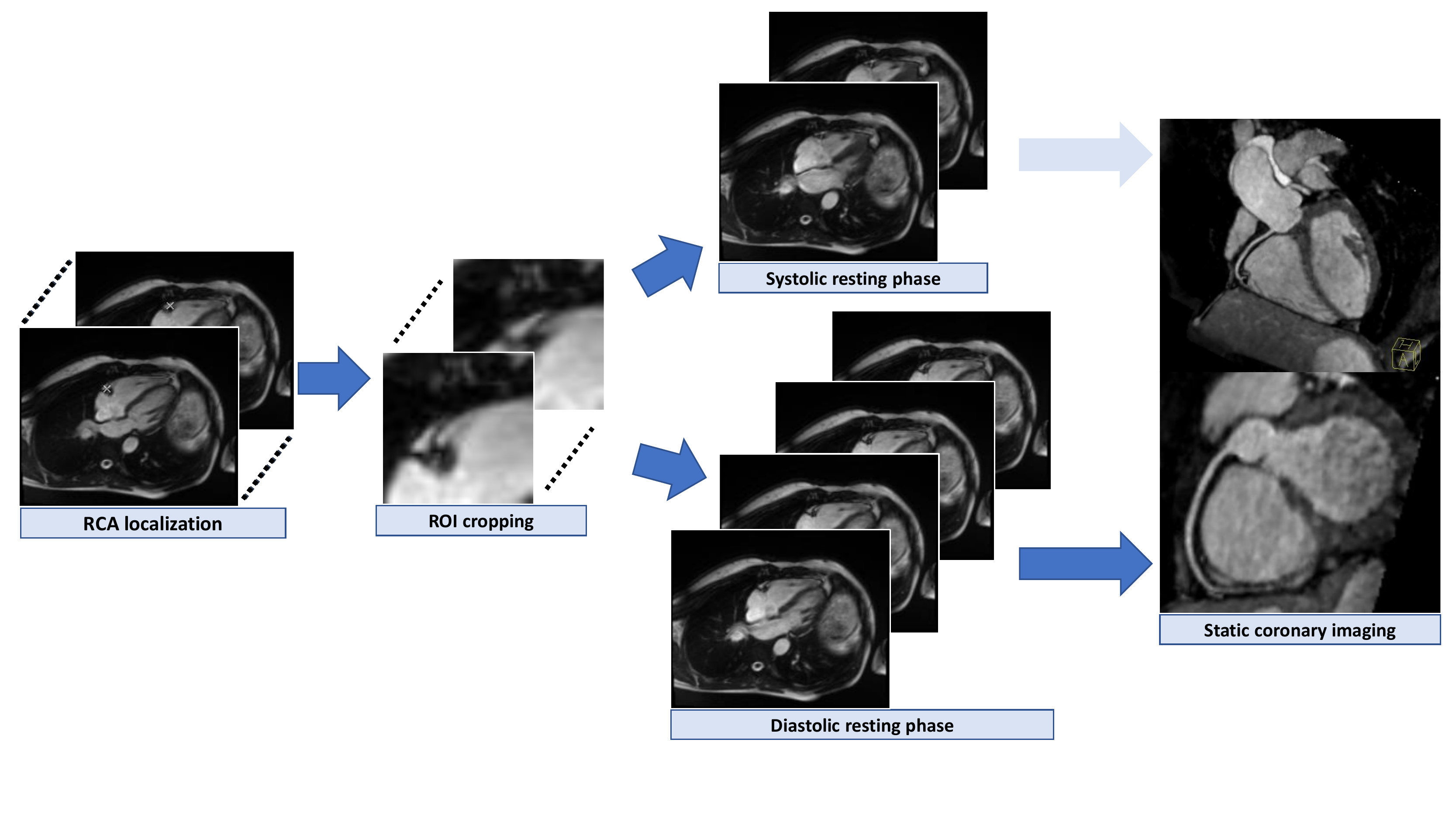}
    		\caption{An example of a volunteer study illustrated with the main steps of the framework pipeline. The outputs are generated directly from the scanner after the proposed system was integrated online. The RCA localization series is the original CINE series with an RCA position marked by a cross. The ROI cropping generated the cropped series based on the RCA localization. From these cropped series, the motion is quantified, from which the RPs are determined. The series that represent the systolic and diastolic resting phase are generated as well. Here, the diastolic resting phase window (dark blue arrow) is applied for the static coronary imaging.}
    		\label{fig:7}
	    \end{figure}

        The detailed results about the performance of the predicted systolic and diastolic RP on the study datasets are listed in Table \ref{table:results_inRP_inFrame}, Table \ref{table:results_acc}. The Bland-Altmann plots showing the performance of start and end detected time point for each systolic and diastolic RP is shown in Figure \ref{fig:5}.
        $MAE_{start,end\ windo{w/frame}_{sys}}$ and $MAE_{start,end\ windo{w/frame}_{dia}}$ for 1.5T datasets was 11.8 ± 16.5 ms (0.38 ± 0.53 frame) and 14.2 ± 18.9 ms (0.48 ± 0.63 frame) for 3T datasets. By using the selected $\tau$, the proposed system resulted in 93.4\% accuracy, sensitivity at 90.1\% and specificity 96.8\% for 1.5T and 92.6\%, 90.7\% and 94.5\% for 3T datasets.
        $MAE_{start,end\ windo{w/frame}_{sys}}$ and $MAE_{start,end\ windo{w/frame}_{dia}}$ was 13.6 ± 18.6 ms (0.46 ± 0.62 frame) when using the datasets independent from field strength for analysis.  The accuracy, sensitivity and specificity achieved by the defined absolute $\tau$, \textless 0.2, was 92.7\%, 90.5\% and 95.0\%.  The automatically classified RPs resulted in a mean max error of ~30 ms, meaning that it deviates by roughly one frame.
        The datasets with RPs with less than 30 ms were discarded (n=9). Further, there was no systolic RP annotated by the expert in 14 cases, and no diastolic RP in 12 cases. These phases were excluded from the analysis. 
        The ${{\widehat{RP}}_{start\ window_{sys}}}$ matched with the annotation, or was off one frame in 93.6\%, the ${{\widehat{RP}}_{end\ window_{sys}}}$ in 97.5\%, the ${{\widehat{RP}}_{start\ window_{dia}}}$ in 93.8\% and ${{\widehat{RP}}_{end\ window_{dia}}}$ in 96.3\%, respectively.
        ${{\widehat{RP}}_{start\ window_{sys}}}$ was detected earlier/later than the expert’s annotation in 27.8\%/5\% and ${{\widehat{RP}}_{start\ window_{dia}}}$ in 20.9\%/16.0\%. ${{\widehat{RP}}_{end\ window_{sys}}}$ was detected earlier/later than the expert’s annotation in 11.4\%/29.1\% and ${{\widehat{RP}}_{end\ window_{end}}}$ in 15.2\%/14.8\%. In average, the ${\widehat{RP}}_{start\ window}$ was selected earlier than the ground truth in 24.3\% of the cases and later in 10.6\%. In 10 cases, outliers were present, off by 2 or more frames. 
        In 10 cases, there were outliers present for the ${\widehat{RP}}_{start\ window}$, off by 2 or more frames. For the ${\widehat{RP}}_{end\ frame}$, it was off by 2 or more frames in 5 cases. \\
        The range of annotated systolic RP was 61.1 ± 24.1 ms and the predicted range of systolic RP was 75.5 ± 32.9 ms. Further the range of annotated diastolic RP was 156.0 ± 102.1 ms and the predicted range was 158.2 ± 104.3 ms.
       \begin{figure}[t]
    		\centering
    		\includegraphics[width=1\linewidth]{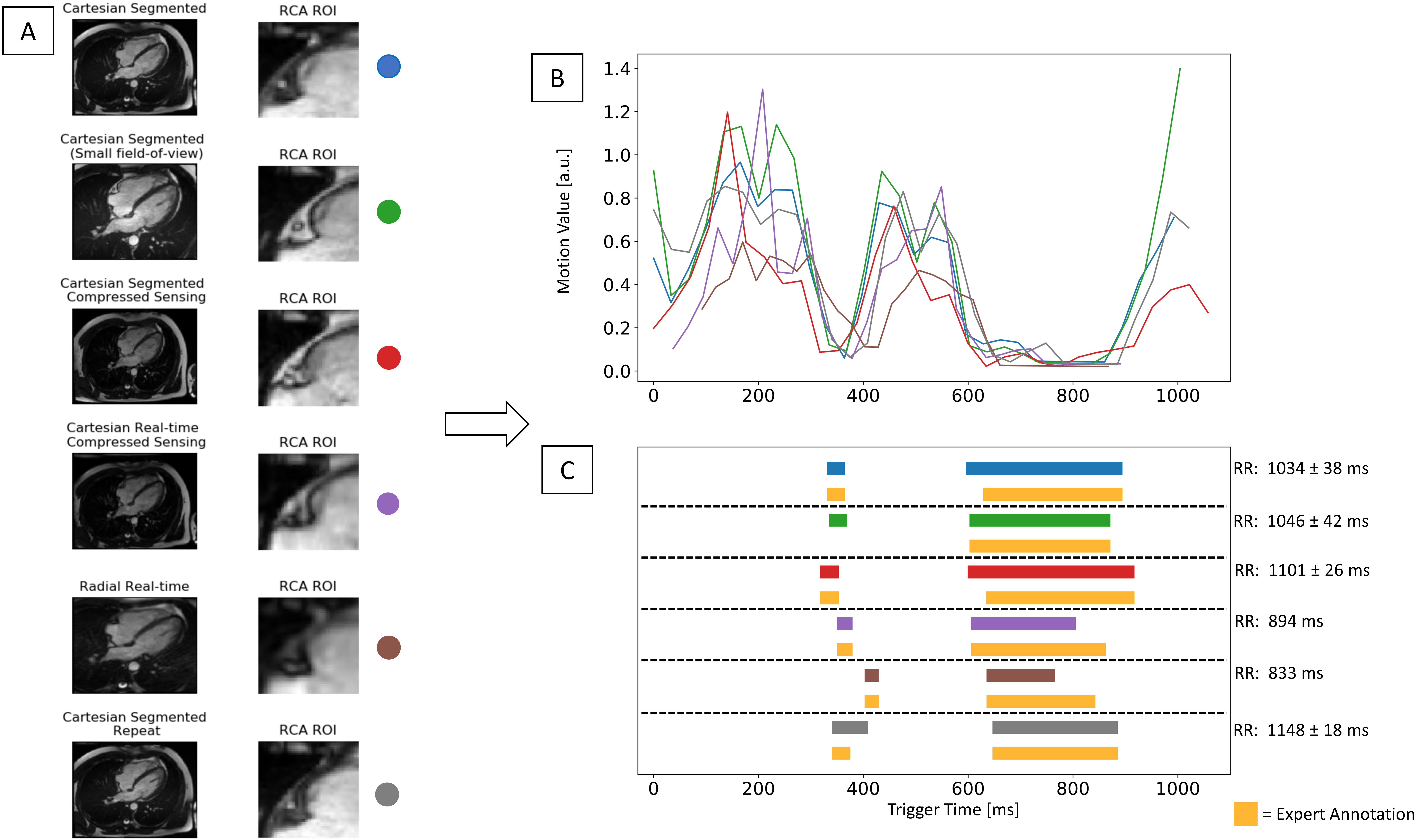}
    		\caption{An example of a study case with 6 different CINE sequences. On the left (A), the first frame of each CINE series and the output of the cropped series are shown. Each color represents one sequence acquisition. On the top right (B), the quantified motion values of each RCA cropped series are plotted over the time points. On the bottom right (C), the classified RPs of each are shown with the expert’s annotation, which is marked in orange color. The RR interval of the real-time CINE acquisitions varied from the segmented CINE, which caused early detection of the RPs.}
    		\label{fig:6}
	    \end{figure}
        In Figure \ref{fig:6}, the robustness of the proposed system is shown in which the system was tested in different sequences including a rescan from a single volunteer. The visualized different CINE outputs are acquired in the order from top to bottom and with ~5min between the first and last acquisition. The predicted start and end systolic phases were matched with annotation in most cases, except in one CINE sequence, where the systolic RP was detected by the system but not by the expert, and in a repeat scan, the end time point was off one frame. The start diastolic RP was detected one frame earlier in 2 cases, and end diastolic RP was detected two frames off in real-time sequences. 
        In an example case, the automatically detected RPs were used for the later 3-D static cardiac acquisition targeted to the RCA. The 3-D RCA visualization with the automatically classified RPs showed no residual motion artifacts (Figure \ref{fig:7}). The computation time of the proposed system was averaged 1.5 seconds.

\section{Discussion}
    The detection of the RCA ROI is successfully and robustly performed by the 3-D DenseNet on the testing dataset and on the study cases. The 3-D based Conv networks leverage the spatial and temporal information from the time-resolved input, rather than learning only spatial information per time point. The size of the fixed ROI ($50 \times 50$mm$^2$) was sufficient in all cases for depicting the RCA at each time point. Further, the network was robustly performed on diverse oriented cases (Figure \ref{fig:oblique}) and different CINE sequences (Figure \ref{fig:6}) which allows the proposed system to be integrated into different clinical protocols. \\
    In terms of quantifying motion values, the approach of taking the Euclidean distance from the predicted RCA pixel coordinate over cardiac phases highly relies on the performance of the network, furthermore, taking the pixel distance measurement for the displacement metric of the anatomy-of-interest between the consecutive frames was not accurate as shown in Table \ref{appendix:table2}. The approaches deriving from the motion values by taking the deformation fields defined by the elastic image registration show a clear advantage, from which the highest accuracy was the one using the weighted deformation fields (see in Table \ref{appendix:table2}). The approach with deformation fields is clearly more robust to slight inaccuracies of localization results. The Gaussian weighting further improves the performance of the system, as it allows to focus on the target-of-interest, and eliminates the area which is not of interest, such as the blood flow in the atrium contained in the RCA ROI. \\
    The metric for assessing the motion values from the weighted deformation fields was reasonably chosen as \nth{50} percentile by the accuracy analysis. The absolute threshold value is selected based on AUC-ROC analysis, evaluated on the testing datasets from the different 1.5T, 3T scanners and CINE sequences and no specific data selection, thus shows versatility in the results (Figure \ref{fig:4}). The classification of the RP by the absolute threshold value is possible due to the quantitative outputs of the deformation field-based approach which enables the detection of the phases with minimal motion, which can be either end-systolic or mid-diastolic or end-systolic and mid-diastolic RPs. 
    In Figure \ref{fig:4}, the quantified motion values of the clinically acquired dataset with different CINE sequences and the corresponding predicted RPs were well matched with the expert’s annotation. As shown in Table \ref{table:results_inRP_inFrame}, Table \ref{table:results_acc}, the proposed framework performed robust in different field strengths as well. Interestingly, in the dataset visualized in the center, there was no RP found by a medical expert, and the proposed system was also able to classify the non-RP case demonstrating the advantage of the system. In such cases, the system gives the user the possibility to take the quantified curve as a reference and select the phase with minimum motion based on the motion curve. \\
    Based on the Bland-Altmann plots (Figure \ref{fig:5}), the systolic RP did not perform as well as the diastolic RP, especially the classification of the start systolic RP was challenging. The predicted systolic RP window was usually slightly longer than the experts’ annotations. However, the mean range difference was 15ms, which is negligible since the temporal resolution of CINE series is ~30ms. In a recent study, the proposed system was evaluated in a clinical validation, in which the interobserver variability was extensively validated and revealed that the automatically detected RPs were consistent with those of the experts \citep{ogawa2022neural}.
    As shown in Figure \ref{fig:7}, the RCA was sharply visualized without severe residual motion artifacts that was acquired during the automatically detected RPs.
    
    The study presents some limitations. First, the proposed method depends on two factors: the performance of the RCA localization network and the registration algorithm that may lead to inaccurate results in the presence of severe artifacts in reconstructed CINE series, even though the network performed well on difficult cases as shown in Figure \ref{fig:oblique} and Figure \ref{fig:worst_cases}. Second, the absolute thresholding used for classification, can be further investigated, whether other algorithms can be used for binary classification. Third, in this study, the focus was highly on method evaluation on patient dataset, however the clinical feasibility and interobserver variability study using this method was performed in \citep{ogawa2022neural}. In the following research, a further clinical study can be of interest to compare the image quality between once acquired with the automatic approach and the other with manually determined resting phases.
    
    Previous methods have shown the feasibility to detect the imaging acquisition window automatically using the shim box volume positioning and cross-correlation calculation \citep{jahnke2005new, ustun2007automated}. However, these semi-automated methods still require the careful positioning of the shim volume coverage. Further, the RP detection is targeted to the whole heart instead of a specific anatomy of interest. These methods were validated on healthy in vivo subjects on a single field strength.
    Moreover, approaches based on the standard deviation of pixel intensity \citep{huang2014automatic} or difference of gradient magnitudes \citep{piccini2017automated} were introduced, allowing the RP detection in real time. This method however performs the RP detection globally based on the entire field-of-view, and the detected RPs are always two RPs as the search is done by two local minima.
    Several approaches have been proposed for the automated determination of targeted RP on regions such as RCA \citep{sato2009approach, asou2018automated}.  A template matching was performed for finding the area with the outer edge of the cross-section of the coronary artery, however the template was defined based on randomly selected five datasets \citep{sato2009approach}. In a more recent study, the regions were extracted from the high-speed component within CINE series by the frequency domain analysis, however by doing so, the cardiac anatomy structures can be disregarded. The authors stated that this method was validated on healthy volunteers, and uncertain the performance on large dataset especially with high heart rates. 
    In \cite{naledi2020automatic}, the deep learning-based RP detection network is built by combining the CNN and Long short-term memory models taking the CINE series as input, and outputs the binary output which is either a RP or not a RP. Therefore, this method is not quantitative and unclear which cardiac structure is weighted, or whether the network tries to detect global RP. 
\section{Conclusions}
    To our knowledge, this proposed study is the first to present the fully automated localized RP detection framework from a CINE series that was validated with a large dataset with multiple 1.5T and 3T scanners acquired with different CINE protocols, such as with free-breathing or breath-held techniques.
    We investigated the robustness and feasibility of the proposed system for fully automated systolic and diastolic RP detection. The proposed system can improve the workflow efficiency, automation, and standardization of the static cardiac imaging that broaden the applicability towards any static cardiac imaging. The RP detection system can be applied in various applications, such as 2-D and 3-D LGE, mapping, 3-D coronary imaging, or any other applications in which the information of RP of heart can be useful. 
    Future work will focus on clinical validations, improving the accuracy of RP classification and integration of automatic detection of other regions, such as the atria and ventricles. 


\acks{This work was supported by Siemens Healthineers GmbH.}

%
\ethics{The local ethics committee approved the study, and a written informed consent was provided by all subjects. The declaration of consent to participate in the study was signed by all subjects.}

\coi{The authors declare no conflict of interest and no financial disclosure in relation to this study. A. Maier is Associate Editor of the Journal of Machine Learning for Biomedical Imaging (MELBA). Seung Su Yoon, Elisabeth Preuhs, Michaela Schmidt, Christoph Forman, Teodora Chitiboi, Puneet Sharma, Juliano L. Fernandes, and Jens Wetzl are employees of Siemens Healthcare GmbH.}

\bibliography{melba-sample}


\clearpage

\end{document}